# Real-Time Stress Monitoring, Detection, and Management in College Students: A Wearable Technology and Machine-Learning Approach

Alan Ta[1], Nilsu Salgin[1], Mustafa Demir[1], Kala Phillips Reindel[2], Ranjana K. Mehta[3], Anthony McDonald[3], Carly McCord[2], and Farzan Sasangohar[1*]

[1]Wm Michael Barnes Department of Industrial and Systems Engineering, Texas A&M University
[2]Telehealth Institute, Texas A&M University
[3] Industrial and Systems Engineering, University of Wisconsin-Madison

***Corresponding Author:** Farzan Sasangohar
Institution: Texas A&M University
PO Box: College Station, TX,
Email address: sasangohar@tamu.edu

**ABSTRACT**

College students are increasingly affected by stress, anxiety, and depression, yet face barriers to traditional mental health care. This study evaluated the efficacy of a mobile health (mHealth) intervention, Mental Health Evaluation and Lookout Program (mHELP), which integrates a smartwatch sensor and machine learning (ML) algorithms for real-time stress detection and self-management. In a 12-week randomized controlled trial ($n$ = 117), participants were assigned to a treatment group using mHELP's full suite of interventions or a control group using the app solely for real-time stress logging and weekly psychological assessments. The primary outcome, "Moments of Stress" (MS), was assessed via physiological and self-reported indicators and analyzed using Generalized Linear Mixed Models (GLMM) approaches. Similarly, secondary outcomes of psychological assessments, including the Generalized Anxiety Disorder-7 (GAD-7) for anxiety, the Patient Health Questionnaire (PHQ-8) for depression, and the Perceived Stress Scale (PSS), were also analyzed via GLMM. The finding of the objective measure, MS, indicates a substantial decrease in MS among the treatment group compared to the control group, while no notable between-group differences were observed in subjective scores of anxiety (GAD-7), depression (PHQ-8), or stress (PSS). However, the treatment group exhibited a clinically meaningful decline in GAD-7 and PSS scores. These findings underscore the potential of wearable-enabled mHealth tools to reduce acute stress in college populations and highlight the need for extended interventions and tailored features to address chronic symptoms like depression.

# INTRODUCTION

## Mental Health Challenges Among College Students

Various studies have highlighted the prevalence of mental health issues among college students [1], [2]. Students enrolled in universities generally face novel stressors, such as the transition from high school to college, the competitive academic environment, as well as the freedom to plan their own schedules [3], [4]. Prior to the COVID-19 pandemic, stress was already a challenging issue that college students faced, with 65.7% of college students reporting “overwhelming anxiety” and 58.7% reporting “more than average” or “tremendous” stress in the Spring of 2019 [5]. Following the advent of the COVID-19 pandemic, the percentage of students facing “moderate” or “high” stress increased to 80.9% in the Fall of 2020, emphasizing the rising stress levels among students in the college environment [6].

Stress is a mental and emotional load on a person’s condition. Stressors cause a physiological response from the sympathetic nervous system, leading to changes in heart rate, typically during moments of stress [7]. A moderate to high level of daily stress can be greatly detrimental to college students’ well-being; these students may see their GPA, confidence, social relationships, and overall mental or physical health impacted by their stress [8]. Thus, it is imperative that college students and their support systems manage stress levels well.

The traditional format of mental health care for stress and other mental health conditions is a combination of psychotherapy and medication [9]. Many universities have begun initiatives into mental health maintenance, typically consisting of traditional therapies such as counseling sessions or educational resources. However, these delivery methods of intervention face their challenges, from the social stigma associated with mental health treatment, the desire to be self-reliant, or the lack of education regarding mental health in general [10]. Also, many students struggle to find the time in their schedules to commit to counseling sessions. Consequently, there is a growing trend of mental healthcare delivery through smartphone applications (apps). The ubiquitous nature of smartphones today makes them the platform with the greatest potential for innovation in the mental health space. These mobile smartphone apps, generally called mHealth apps, introduce monitoring and self-management tools for patients and users, who may be more inclined to take the initiative in their mental health [10].

**mHealth: A Digital Approach to Mental Health Management**

Health coaching, especially when delivered digitally via mHealth, has been shown to improve health outcomes in populations with chronic diseases due to its accessibility, time efficiency, and lower cost compared to in-person coaching [11]. mHealth apps have been previously shown to improve not only symptoms of depression and anxiety, showing great promise in self-management of mental health issues, but also physical chronic conditions. Particularly, the majority of mobile apps that have been tested for their effectiveness in changing health behavior have been for physical activity, diet, and drug and alcohol use, as compared to mental health. Furthermore, mixed levels of mHealth app effectiveness on health behaviors or outcomes, prompting the need for further research into this area [12]. Additionally, it can be said that students generally prefer to use self-management techniques for their mental health conditions, meaning mHealth apps have a strong potential to improve the lives of university-age students [13].

Furthermore, advancements in wearable sensors, such as smartwatches, are nonintrusive, accessible, and designed to gather physiological data in real time [14]. When used in tandem with self-management apps, these devices can provide a comprehensive approach and have the potential for more care-integrated interventions to improve stress outcomes in patients and students.

Research indicates that understanding and being open to stress and the body's response to stress can improve symptoms of anxiety and depression [15]. Studies indicate that a clear physiological response to stress affects heart rate and other measurable metrics, and wearable sensors can effectively mitigate stress by combining detection with interventions [16]. With the advent of mHealth apps and wearable sensors, there is a need to improve stress detection methods in naturalistic settings.

There is growing research that supports the efficacy of Digital Mental Health Interventions (DMHIs) in addressing psychological distress and mental health conditions among college students. A recent meta-analysis provided evidence that indicates DMHIs are effective at reducing both anxiety and depression, with medium effect sizes observed across randomized controlled trials [17]. These effects hold across various delivery formats, including fully automated mobile apps and guided web-based programs. Systematic reviews have further highlighted the utility of DMHIs in real-world college settings, emphasizing their reach and implementation potential, particularly when interventions align with student needs for accessibility, flexibility, and evidence-based therapeutic content [18]. Additionally, student-centered research has shown that most young adults not only experience high levels of mental health needs but also express a willingness to engage with digital solutions, especially those that integrate coaching support, crisis tools, and behavioral self-help modules [19]. Despite this promising evidence, more research is needed to evaluate DMHI efficacy in naturalistic contexts over sustained periods.

To ensure that mHealth solutions are not only accessible but also therapeutically valuable, it is important to incorporate features that reflect evidence-based mental health practices. One such practice is expressive writing, commonly implemented through journaling, which mental health professionals often recommend to help individuals process emotions and reflect on their experiences. Research has shown that this type of emotional expression can significantly improve psychological and physical health outcomes [20]. Another essential method to manage stress is slow breathing exercises, another clinically supported strategy that has been shown to reduce symptoms of anxiety, depression, anger, and confusion by modulating central nervous system activity and promoting relaxation [21]. By integrating such features, mHealth apps not only mirror established therapeutic approaches but also empower their users to manage their mental health in a private, accessible, and proactive manner.

It is important to note that while mHealth applications provide a platform that is easily accessible and holds promise to improve mental health conditions, they do not act as complete alternatives to traditional forms of therapy and counseling. There is no strong evidence that supports mHealth apps being more effective than traditional therapy. Rather, while mHealth apps can act as standalone tools, most are designed to complement and enhance traditional therapies [22]. There are several circumstances in which it might be more effective to employ a combination of traditional therapeutics and mHealth apps. For instance, many college students struggle with the short-term nature of traditional therapy, limited to a certain number of sessions, which leads to them feeling as if the continuity of care afterward is poor. In situations like this, implementing a monitoring and self-management tool, such as an mHealth app, alongside traditional therapy can extend the positive effects of psychological counseling and facilitate the application of skills learned in the therapy room to real life.

**Previous Research and Novel Contributions of This Study**

Previous studies into the real-time physiological stress detection have generally been conducted in controlled lab environments. Mishra et al. (2020) defined and detected stress versus non-stressed states for individuals using physiological data collected from wearable sensors within controlled environments [23]. Additionally, Ziyadidegan et al. (2022) report that most stress detection methods introduce forms of bias due to lab-generated stress data and tend to focus on stress resulting from physiological stressors rather than mental stressors [24]. Furthermore, a study into mHealth apps for the mental health of college students has also shown promise in decreasing symptoms of anxiety, depression, and stress, but most available apps lack human factors approaches to make the app more usable and engaging in the long term [25].

The growing prevalence of stress among college students has prompted the development of innovative tools to measure and address this issue. Wearable sensors combined with ML-driven algorithms offer a novel approach to monitoring stress by providing real-time data and predictive insights. However, the extent to which mental health interventions can alter these stress patterns remains unclear. Therefore, the current study introduced in this paper focuses on stress detection and self-management of mental health conditions like anxiety, depression, and general stress via a longitudinal study in a naturalistic setting where stress was monitored continuously via a combination of wearable sensors and a mHealth app. Participants in the study were equipped with the necessary tools to track and manage their stress and mental health without needing to be present in a laboratory environment. Thus, the changes in their mental health conditions are more likely to reflect their day-to-day lives and fluctuations in their personal and academic careers. In addition, this study stands out due to the heavy focus on engineering and human factors involved in the development of the Mental Health Evaluation and Lookout Program (mHELP) app, the central hub for self-management used in this study. The app features modern technology, with the integration of a smartwatch sensor and novel Machine Learning (ML) algorithms that take physiological input from said sensor to predict stressful moments in participants' lives in real-time. Furthermore, several iterations of usability testing were performed in designing the mHELP app to make it more user-friendly and include the most desired features.

# CURRENT STUDY

## Experimental Design

This study examines the impact of mental health interventions on stress management among college students, leveraging data collected during the Fall 2021 mHELP study. The primary objective of this research is to determine whether mental health interventions influence the frequency of stress moments. These moments are identified through participant self-reports and an ML-based algorithm embedded in the mHELP application. Additionally, the study explores differences in stress reports, Heart Rate (HR) patterns, and accelerometer data between the control and treatment groups to uncover potential behavioral and physiological changes across conditions.

Participants were randomly assigned to one of two groups (between-subjects): (1) a control group that used the mHELP application exclusively for stress recording and (2) a treatment group that received access to therapeutic interventions integrated into the app and personal therapy sessions. All the participants installed the mHELP application on their iPhones, enabling seamless data collection from their phones and connected wearable devices. The application operated continuously for three months (12 weeks: within-subjects), recording HR data and other relevant physiological and behavioral metrics. Experimenters reviewed the data daily to ensure consistency and accuracy, maintaining rigorous standards throughout the study.

## Research Questions and Hypotheses

In this study, we addressed two main Research Questions (RQs):

*RQ1:* Do digital mental health interventions impact the frequency of stressed moments in college students over time, measured via wearable sensors, ML algorithms, and self-reports? So, we hypothesized ($H_1$) that extended use of the mHELP application's suite of self-management tools will correlate with a decreased rate of identified stress moments, leading to better mental health conditions. A central hub that is easily accessible to college students should help them identify and manage their stress and stressful moments more efficiently. A tool like mHELP that reports real-time stress moments can give participants greater awareness of their mental condition. When combined with management techniques, this allows students to better react and cope with stressful moments, which tend to act more acutely. In contrast, participants in the control group who lack these self-management tools are predicted to have slight changes from their baseline measurements, with fluctuations in mental health conditions likely due to changes in their personal and academic lives rather than any interventions.

*RQ2:* Are the management techniques provided through mobile health applications significantly and negatively associated with anxiety, depression, and stress in college students over time? Following similar reasoning, it is hypothesized ($H_2$) that extended use of the mHELP application's suite of self-management tools will correlate with better mental health outcomes. The mHELP app acts as a digital coaching and monitoring tool for students' stress, anxiety, and depression. Understanding these symptoms of anxiety, depression, and stress alongside attempting different coping strategies with traditional therapy sessions should lead to better mental health outcomes as students take the initiative to improve their condition. These surveys respond to more chronic mental health conditions, so management techniques will likely need to be used over several months to see positive effects. It is expected that the control group participants will not show strong improvement on self-assessments reporting symptoms of anxiety, depression, and general stress; fluctuations in mental health conditions will likely be due to changes in their personal and academic lives rather than any interventions.

## METHODOLOGY

### Participants

One hundred twenty-five university students enrolled in the study, and of these, 117 successfully completed: 92 were female, 23 were male, one was non-binary, and one was nondisclosed. Eighty participants were undergraduate students, and 37 were graduate students. To be eligible for the study, participants needed to show moderate symptoms of anxiety, indicated by a score of seven or higher on the Generalized Anxiety Disorder 7-item scale (GAD-7). Additionally, they must have completed the online questionnaire before enrollment and own an iPhone. Exclusion criteria included anyone unable to consent.

To achieve sufficient statistical power to evaluate the impact of the mHELP app on students' stress levels, a power analysis was conducted by using G*Power 3.1.9.7 [26] to determine the required sample size for a mixed Analysis of Variance (ANOVA) with a within-between interaction. The analysis was based on an effect size of $f$ = 0.25, an α error probability of 0.05, a power (1 - $\beta$) of 0.95, and 12-week measurement points. To address potential violations of sphericity, a nonsphericity correction (ε) of 0.70 and a correlation among repeated measures of 0.30 were also applied. The results indicated that a minimum of 32 participants (16 in each condition) is required to achieve sufficient power to detect significant differences. To address potential attrition, a larger number of participants was recruited, as studies indicate that up to 50% of individuals diagnosed with mental health issues may drop out [27].

**Procedure**

Eligible students who consented to participate in the study were randomized into two conditions: Students in the experimental group could access the mHELP app's full capabilities throughout the 12 weeks, while the control group could only record stress moments and take psychological surveys during that time. From the pool of students who completed the study, 31 were in the control group, while 86 were in the experimental group. To improve the likelihood of identifying changes within the intervention group while preserving a sizable control group, participants were sequentially randomized into either the intervention or control group using a 3:1 allocation.

Students randomized to the treatment condition were invited to an onboarding session in the academic Fall semester of 2021. At the onboarding session, students were introduced to the goal of the study and were loaned an Apple Watch pre-loaded with the mHELP app. Students were given links to download the mHELP app onto their iPhones for pairing and were taught how to record stressful moments through either their smartwatch or smartphone app. The app uses the capabilities of the wearable Apple Watch sensors to record heart rate and accelerometer data, which is stored in the iPhone app alongside self-reported and suggested stress events, which are determined through a proprietary machine learning model. The app operated non-stop except when the watch was taken off, such as when charging the device or showering. The study began at the end of September 2021, and students were asked to keep the mHELP app running continuously and confirm their stressful moments for 12 weeks. Students, regardless of group placement, were also asked to complete weekly psychological questionnaires to report the overall condition of their anxiety, depression, and general stress symptoms.

### Equipment, Apparatus, and Digital Tools

Each participant was equipped with an Apple Watch (Series 4 or 5) with a preinstalled mHELP app that connected to the app on their iPhone device. Weekly psychological assessment surveys were programmed into the mHELP app. These assessments included the Generalized Anxiety Disorder-7 (GAD-7) for anxiety [28], the Patient Health Questionnaire (PHQ-8) for depression [29], and the Perceived Stress Scale (PSS) for stress [30].

### The mHealth Application

The study used the Mental Health Evaluation and Lookout Program (mHELP) application, a fully developed smartphone and paired smartwatch application that acts as a central hub for mental health symptom monitoring and management. An iterative design process was used to map out the features of the application. Several rounds of usability testing, individual interviews, and focus groups were performed to ensure the most desirable, effective, and engaging features were implanted into the app layout. The mHELP smartphone application includes various features, including self-management activities, a digital coaching plan, external links, and several health-tracking capabilities, as seen in Figure 1. The home page consists of a calendar with daily tasks, such as completing a survey or reading an educational article. Aside from the home page, three other screens can be accessed from the navigation bar at the bottom of the screen. The "Engage" screen includes a suite of self-management features, such as relaxation media, breathing exercises, and educational content. The "Link" screen provides several contacts and resources, including on-demand bookings for counseling via a telehealth platform staffed by appropriate mental health professionals, and emergency resources, allowing users to easily add emergency contacts and reach out to emergency hotlines. Through the telehealth platform, students were eligible for up to 5 telehealth counseling sessions, although they were not required to use all of them. The "Progress" screen includes different health tracking capabilities, such as stress moments, heart rate, sleep, and medication.

Additionally, the mHELP smartwatch app, as depicted in Figure 2, is paired with the smartphone app and acts as a wearable sensor to securely collect and transfer physiological data to the smartphone app. The smartwatch app allows for recording stressful moments, as detected by the app's machine learning algorithm. Furthermore, the watch app allows the user to track heart rate data in real-time and also provides some of the capabilities of the smartphone app, such as breathing exercises and access to counseling services at the click of a button. During real-time stressful moments, the app detects high stress and then can prompt users to perform a breathing or focus exercise, pulling them out of a panicked or high anxiety state. Overall, the mHELP app has a variety of features for users to engage with, tailored towards empowering students to self-manage their stress and mental condition.

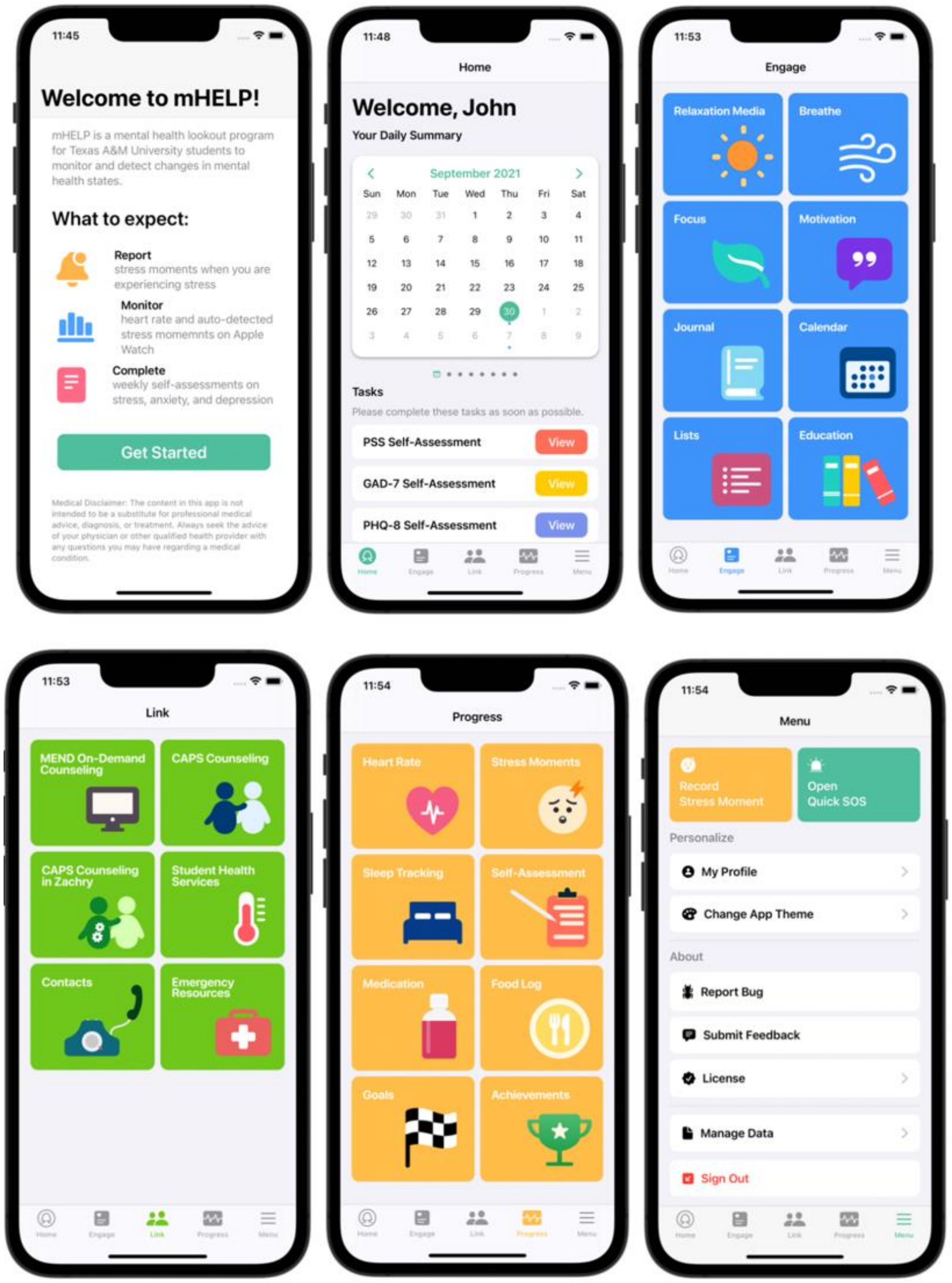

**Figure 1:** Sequence of core mHELP smartphone navigation screens. From left to right, the app displays (1) a welcome and onboarding screen, (2) the home dashboard with calendar and assessment tasks, (3) the Engage tab offering stress management tools (e.g., breathing, journaling, motivation), (4) the Link tab with counseling and emergency resources, (5) the Progress tab tracking health metrics (e.g., heart rate, stress moments, sleep), and (6) the Menu tab for personalization, data management, and quick access to stress reporting.

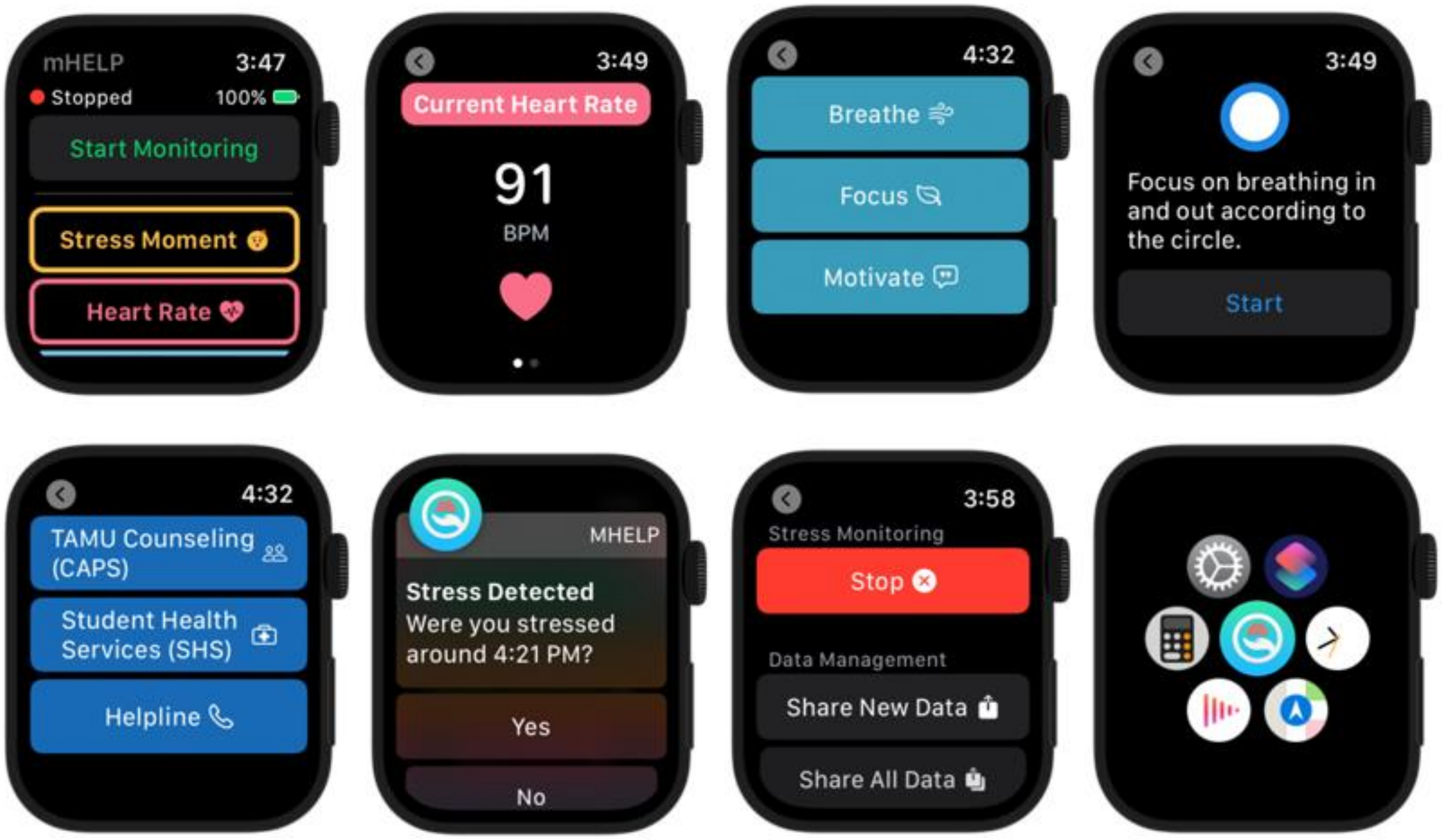

**Figure 2.** Sequence of mHELP Apple Watch interface screens. The smartwatch interface allows users to (1) start/stop monitoring, report stress moments, and view heart rate, (2) access real-time heart rate readings, (3) engage in brief interventions such as breathing, focus, and motivation exercises, (4) follow guided breathing prompts, (5) quickly reach counseling and health services, (6) respond to detected stress prompts, (7) manage data sharing settings, and (8) navigate among installed watch apps.

## Measures and Metrics

The primary dependent variable, "Moments of Stress" (MS), is a conceptual real-time measure derived from a combination of physiological heart rate (beats per minute - bpm) measurements and subjective self-reports, both collected through the mHELP application on the Apple Watch. Participants were prompted by the app at random intervals and after periods of elevated heart rate to provide momentary self-assessments of stress using a single-item scale (e.g., “Are you feeling stressed right now?”), enabling an acute, context-sensitive capture of perceived stress.

Continuous HR data were recorded using either the Apple Watch Series 4 or 5, measuring bpm at a frequency of 1 Hz. [31], enabling high-resolution monitoring for precise identification of physiological stress indicators. Stress events were detected using an integrated algorithm analyzing HR fluctuations, and these were automatically logged in the dataset under a designated “stress detection” column. The mHELP application also allowed participants to self-report significant stress moments by tapping on their watch faces (Sadeghi, McDonald et al., 2022; Sadeghi, Sasangohar et al., 2022). Binary coding classified events, with a score of ‘1’ assigned to stress moments detected by the algorithm and subsequently approved by the user, and ‘0’ representing moments without detected or self-reported stress. Participants’ self-reported stress moments were timestamped and recorded to complement sensor-driven detection, providing additional context. Both algorithm-detected and self-reported stress events were aggregated daily to derive individual stress scores, reflecting the total frequency of MS for each participant.

Three subjective measures of stress and psychological functioning – the GAD-7, PHQ-8, and PSS – were collected as secondary outcome measures. Once a week, the mHELP app would prompt the students to take these psychological surveys. If students were unresponsive or behind on their tasks, the app would send a daily notification with a reminder to complete these surveys. The GAD-7 survey is a brief questionnaire that asks individuals how often, within the last two weeks, they experience various symptoms of anxiety from a range of “not at all” to “nearly every day” [28]. The PHQ-8 is a similarly designed survey that asks individuals about symptoms of depression [29]. Both assessments are designed based on the criteria listed for the respective condition in the Diagnostic and Statistical Manual of Mental Disorders, Fourth Edition (DSM-IV). The PSS items evaluate the degree to which individuals believe their life has been unpredictable, uncontrollable, and overloaded during the previous month [30]. A higher survey score on each of the three scales indicated a higher severity of the corresponding mental health condition.

## DATA ANALYTICS AND RESULTS

### Moments of Stress (MS)

Prior to the analysis, the normality of the response variables was assessed using the Shapiro-Wilk test, which revealed significant deviations from normality for MS ($W = 0.699$, $p$ <.001), as depicted in Figure 3 (left). To address this issue, a Box-Cox transformation was applied to the MS variable. However, the subsequent Shapiro-Wilk test still indicated a violation of normality ($W = 0.912$, $p < .001$), as shown in Figure 3 (right). Although the distribution showed some improvement, it remained non-normal, which may be attributed to the binary nature of the data that inherently resists transformation to a normal distribution. As depicted in Figure 3 (right), after the Box-Cox transformation, the scale of the x-axis changes significantly due to the transformation process, which alters the numerical range to make the data more normally distributed. Given the positively skewed nature of the continuous response variable and accounting for the longitudinal structure of the experimental design, a Generalized Linear Mixed Model (GLMM) was applied to predict MS. A Gamma distribution with a log-link function was used to address the skewed distribution, as it is well-suited for positively skewed data and ensures that predicted values remain strictly positive [34].

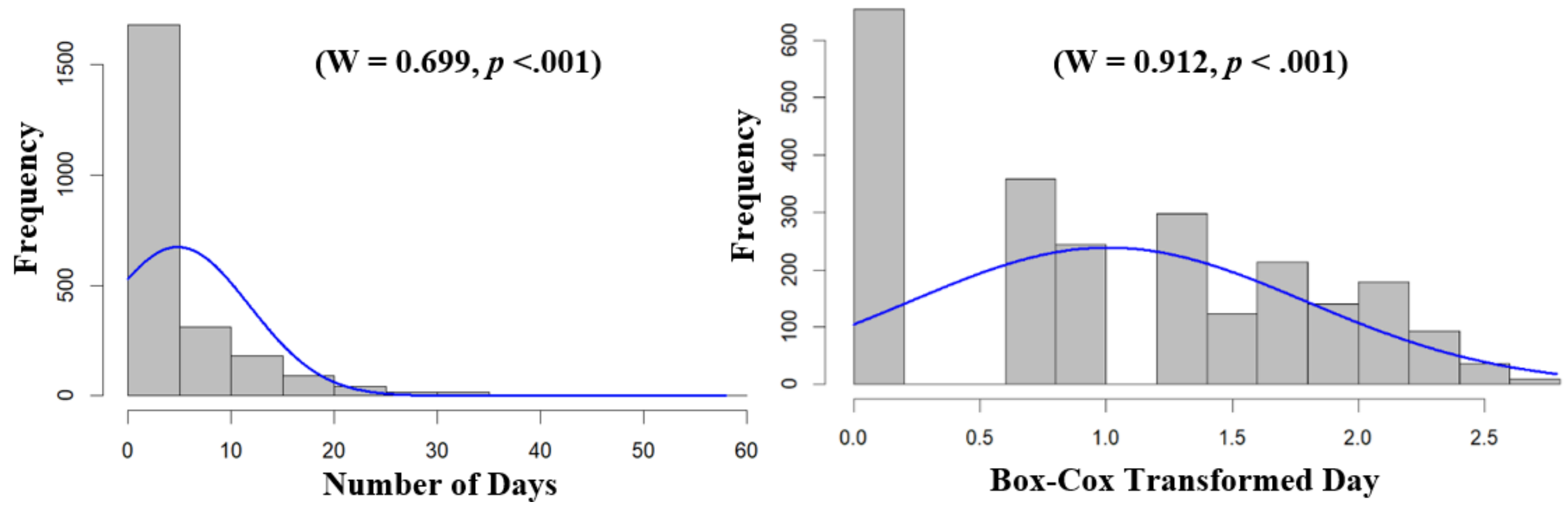

**Figure 3.** Histograms of moments of stress across the number of days: **Left:** Original moments of stress, **Right:** Box-Cox transformed moments of stress.

GLMM was fitted using maximum likelihood estimation with Laplace Approximation [35]. Firstly, to address potential issues arising from zero '0' values in the dataset, a small constant was added to the response variable, creating a shifted version of $MS_{shift}$. This adjustment ensures compatibility with the Gamma distribution used in the GLMM, as the Gamma family requires strictly positive values for analysis. By applying this transformation, the integrity of the statistical assumptions underlying the model is maintained while preserving the meaningful variability in the original data. The model was fitted using the `bobyqa` optimizer [36] with a maximum of $1 \, X \, 10^5$ iterations to ensure convergence. Analyses were conducted using the `glmer()` function from the `lme4` package [37] in RStudio.

To predict $MS_{shift}$, the model included fixed effects as condition (i.e., control vs. treatment) and Time (Linear). At the start of the model-building steps, both linear and higher-order polynomial terms of time, i.e., quadratic and cubic terms, were initially considered; however, these higher-order terms were ultimately excluded from the model because including them did not significantly improve the model's fit or provide additional explanation for variance in $MS_{shift}$ ($p > .05$). Additionally, random intercepts and slopes for Time (Linear) were included for each participant, accounting for individual variability in both baseline fixation duration and changes over time. Formula 1 includes fixed and random effects:

$$\log(E[\hat{Y}_{ij} \mid X_{ij}]) = \beta_0 + \beta_1 c_{ij} + \beta_2 t_{ij} + \beta_3 (c_{ij} \cdot t_{ij}) + u_{0j} + u_{1j} c_{ij} + \varepsilon_{ij} \quad (1)$$

In Formula 1, the term $E[MS_{Shift,ij} \mid X_{ij}]$ represents the expected value (mean) of $MS_{Shift,ij}$ for observation $i$ of subject $j$, given the set of predictors ($X_{ij}$), which includes the fixed effects of: intercept ($\beta_0$) – captures the baseline log-transformed response when all predictors are zero, the condition ($\beta_1 c_{ij}$; i.e., control vs. treatment) on the response, the linear time effect ($\beta_2 t_{ij}$) – models how the response evolves over time, and the interaction term $\beta_3(c_{ij} \cdot t_{ij})$ – indicates how the effect of time differs across conditions. In addition, the model incorporates a random effects component ($u_{0j}+u_{1j}c_{ij}$) to account for subject-specific deviations from the fixed effects, including a random intercept ($u_{0j}$), capturing variability in the baseline response for each subject $j$, and a random slope ($u_{1j}$), allowing the effect of the condition ($c_{ij}$) to vary across subjects. Both $u_{0j}$_and $u_{1j}$ are assumed to follow a multivariate normal distribution with a mean of zero and a variance-covariance structure $N(0, \Sigma)$. $\varepsilon_{ij} \sim N$ represents the residual variance. Results were interpreted within this interaction framework to capture insights into how time effects vary across combinations of conditions and time (see Table 2).

The random effects structure of the model included varying intercepts and slopes across participants to account for individual differences in baseline values and treatment effects. The intercept variance ($\sigma^2$ = 1.19, *SD* = 1.09) indicates moderate variability in participants' baseline levels of the dependent variable $MS_{Shift}$. The variance in the treatment slope ($\sigma^2$ = 2.04, *SD* = 1.43) suggests substantial individual differences in how participants responded to the treatment condition over time. The correlation between the intercept and the treatment slope ($r$ = 0.01) is close to zero, indicating that the initial levels of $MS_{Shift}$ and the treatment effect were largely independent of each other. The residual variance ($\sigma^2$ = 2.04, *SD* = 1.43) reflects the within-participant variability unexplained by the model.

The fixed effect results are summarized in Table 1, which presents the estimates of unstandardized coefficients ($\beta$) with Standard Errors (SE), standardized coefficients ($\beta_{std}$), *t*-values, and *p*-values. Accordingly, all the fixed effects were statistically significant, $p < .001$, and expressed in Formula 2 as an exponentiated version (i.e., back-transformed). Based on the significant interaction effect (treatment by time), the simple slope analysis indicates that the rate of decline in the moment of stress is significantly steeper for the treatment group [$\beta_{Std} = -0.127$ (0.013), $t(81) = -9.74$, $p < .001$] than the control [$\beta_{Std} = -0.026$ (0.008), $t(81) = -3.29$, $p < .001$], demonstrating that the treatment effect becomes more pronounced over time, see Figure 4. For the treatment effect**,** the coefficient indicates a statistically significant negative impact of the treatment condition compared to the control, regardless of time, suggesting that, on average, the log-transformed moment of stress was lower in the treatment condition, reflecting a reduced response relative to the baseline. Similarly, the linear time effect indicates a significant decline in the response over time, regardless of condition, underlining that the log-transformed outcome shows a linear decrease as time progresses. In summary, these results indicate that the treatment condition starts at a lower baseline and shows a steeper decline in response over time compared to the control, supporting the conclusion that the treatment's impact becomes increasingly negative as time progresses.

$$\hat{Y}_{ij} = \exp(0.018 - 0.024 \cdot c_{ij} - 0.026 \cdot t_{ij} - 0.102 \cdot (c_{ij} \cdot t_{ij})) + \varepsilon_{ij}, \quad \varepsilon_{ij} \sim \mathcal{N}(0, 2.035) \tag{2}$$

**Table 1.** GLMM fixed effects results for MS.

| Measure/ Formula | Term | *B* (SE) | $\beta_{Std}$ | t-value | Pr(>\|z\|) |
|---|---|---|---|---|---|
| **MS/ Formula 2** | Treatment | -1.61 (0.25) | -0.02 | -6.34 | <.001 |

| | | | | |
|---|---|---|---|---|
| Time (Linear) | -1.72 (0.52) | -0.03 | -3.29 | <.001 |
| Treatment * Time (Linear) | -6.80 (0.70) | -0.10 | -9.69 | <.001 |

**Note.** "$\beta$" and "$\beta$ (*SE*)" refer to the unstandardized regression coefficient and its Standard Error, respectively, while "$\beta$" denotes the standardized regression coefficient.

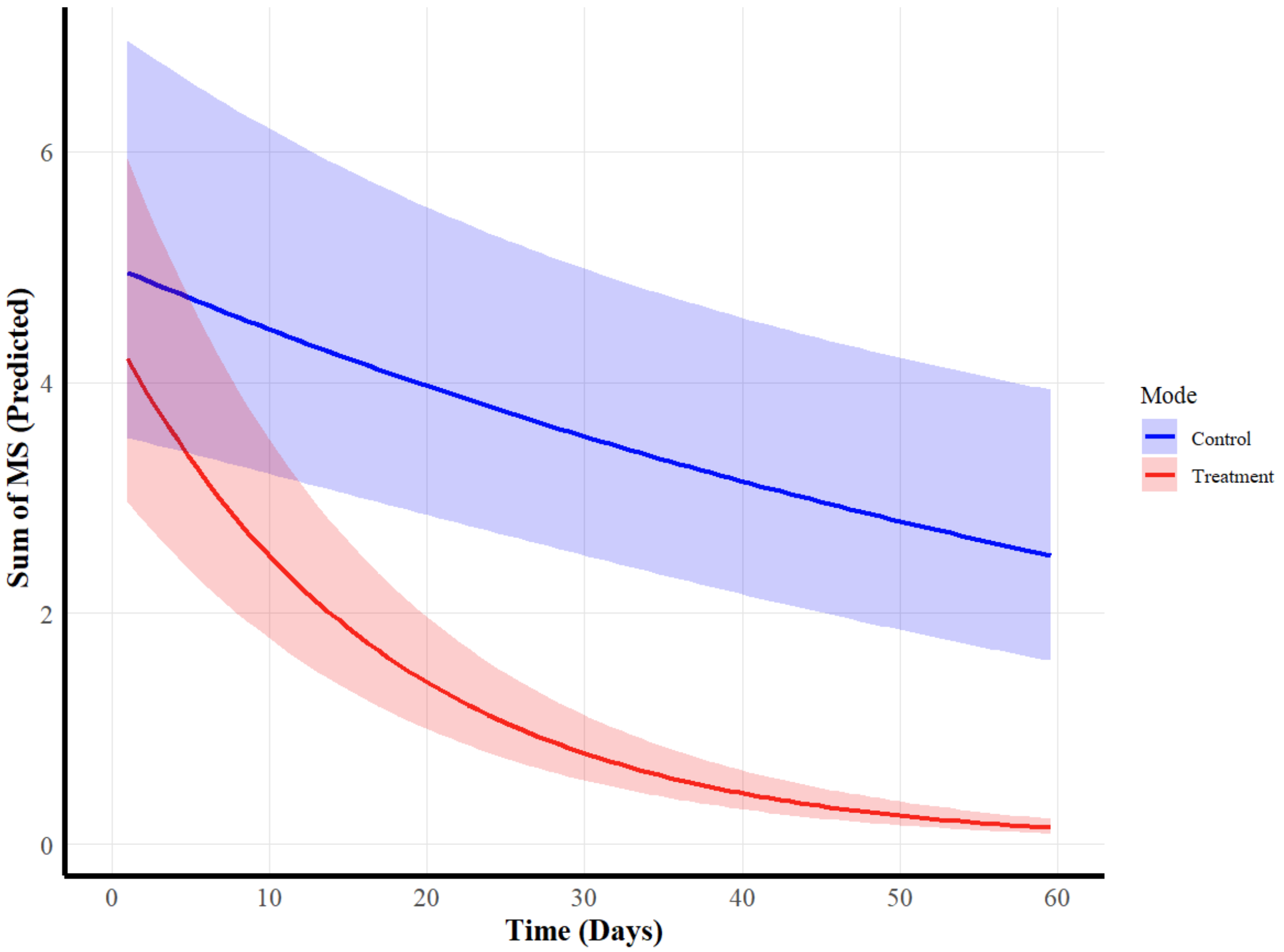

**Figure 4.** Predicted MS over time (linear) across the conditions. Shaded areas indicate 95% Confidence Intervals (%95 CI).

**Self-Reported Measures: GAD-7, PHQ-8, and PSS**

Prior to the analysis, the normality of the response variables of GAD-7, PHQ-8, and PSS was assessed using the Shapiro-Wilk test, which revealed significant deviations from normality for all three measures: GAD-7: $W = 0.976$, $p < .001$; PHQ-8: $W = 0.980$, $p < .001$; PSS: $W = 0.981$, $p < .001$. To address this issue, a Box-Cox transformation was applied to each of these three variables; however, the subsequent Shapiro-Wilk test still indicated a violation of normality for all three (GAD-7: $W = 0.984$, $p < .001$; PHQ-8: $W = 0.987$, $p < .001$; PSS: $W = 0.990$, $p < .001$). Given the positively skewed nature of the continuous response variables and accounting for the longitudinal structure of the experimental design, a GLMM was applied to predict each of the three measures with a Gamma distribution with a log-link function for positively skewed data. Therefore, Formula 3 was used across each self-reported measure.

$$\log(E[\hat{Y}_{ij} \mid X_{ij}]) = \beta_0 + \beta_1 c_{ij} + \beta_2 t_{ij} + \beta_3 (c_{ij} \cdot t_{ij}) + u_{0j} + \varepsilon_{ij} \quad (3)$$

According to the self-reported measure's GLMM, the GAD7 model's random effects structure included varying intercepts for participants (*N* = 125) to address individual differences in baseline values of the dependent variable, with intercept variance ($\sigma^2$ = 0.067, *SD* = 0.259) indicating moderate variability in baseline stress levels. Random slopes for time or condition were excluded due to convergence issues; specifically, adding a random slope for condition led to non-convergence, while a random slope for linear time resulted in numerical instability. To maintain model stability and interpretability, a simpler model with random intercepts only was utilized, which was also applied to the PSS and PHQ8 models. In the PSS model for participants (*N* = 126), the intercept variance ($\sigma^2$ = 0.018, *SD* = 0.134) reflected low variability in baseline stress levels among individuals, with residual variance ($\sigma^2$ = 0.085, *SD* = 0.291) capturing unexplained within-participant variability. For the PHQ8 model for participants (*N* = 125), the intercept variance ($\sigma^2$ = 0.134, *SD* = 0.366) indicated moderate variability in baseline depression scores. In contrast, the residual variance ($\sigma^2$ = 0.266, *SD* = 0.516) accounted for within-participant variability not explained by fixed effects.

Table 2 summarizes the findings for each of the measures' fixed effects. Accordingly, only the main effects of the linear time were statistically significant for GAD7 and PSS (*p* = 0.009 and *p* = 0.004, respectively). That is, as time progressed, anxiety and stress decreased significantly, which are enacted in Formulas 4 and 5 as exponentiated versions (i.e., back-transformed), but depression remained stable (PHQ8). Although the interaction effects were insignificant for the three measures (i.e., no evidence that the slopes differ), visualizing the simple slopes of each finding in Figure 5 revealed negative trends, particularly for GAD7 and PSS, but not for PHQ8, which may demand simple slope analyses. Therefore, we still conducted the simple slope analysis for each self-reported measure.

**Table 2.** GLMM fixed effects results for GAD7, PSS, and PHQ8.

| Measure/ Formula | Term | $\beta$ (SE) | $\beta_{Std}$ | *t*-value | *Pr*(>\|*z*\|) |
|---|---|---|---|---|---|
| **GAD7 / Formula 4** | Treatment | -0.07 (0.11) | -0.004 | -0.62 | 0.532 |
| | Time (Linear) | -0.45 (0.17) | -0.025 | -2.58 | 0.009 |
| | Treatment * Time (Linear) | -0.22 (0.21) | -0.012 | -1.05 | 0.293 |
| **PSS / Formula 5** | Treatment | -0.004 (0.06) | -0.0002 | -0.06 | 0.953 |
| | Time (Linear) | -0.28 (0.10) | -0.012 | -2.9 | 0.004 |
| | Treatment * Time (Linear) | -0.08 (0.11) | -0.004 | -0.71 | 0.475 |
| **PHQ8** | Treatment | 0.02 (0.15) | 0.003 | 0.12 | 0.902 |
| | Time (Linear) | -0.11 (0.06) | -0.019 | -1.86 | 0.063 |
| | Treatment * Time (Linear) | 0.002 (0.07) | -0.0003 | -0.02 | 0.983 |

**Note.** "$\beta$" and "$\beta$ (*SE*)" refer to the unstandardized regression coefficient and its Standard Error, respectively, while "$\beta_{Std}$" denotes the standardized regression coefficient.

The simple slope analysis for the GAD7 indicates that, over time, the treatment group [$\beta_{Std} = -0.037$ (0.015), $t(865) = -2.46$, $p < .014$] experienced a steeper decline in symptoms of anxiety compared to the control group [$\beta_{Std} = -0.025$ (0.009), $t(865) = -2.58$, $p = .009$], see Figure 5(a). Yet, this difference in the decline of the slopes was not statistically significant between the conditions due to an insignificant interaction effect ($p = .293$; see Table 2). Although the variance in the control group is quite large, which may explain why the difference is not significant. This issue could potentially be addressed with a larger control dataset in the future.

Similarly, for PSS, although there was an insignificant interaction effect of the condition by time ($p = .475$), the treatment group [$\beta_{Std} = -0.016$ (0.007), $t(885) = -2.41$, $p < .016$] experienced a steeper decline in self-reported stress compared to the control group [$\beta_{Std} = -0.012$ (0.004), $t(885) = -2.90$, $p = .004$]; yet both of these conditions' decrement of stress symptoms were statistically significant, as illustrated in Figure 5 (b). For PHQ8, simple slope analysis reveals that there were no significant decrements in both the treatment [$\beta_{Std} = -0.020$ (0.016), $t(867) = -1.21$, $p < .228$] and control conditions [$\beta_{Std} = -0.019$ (0.011), $t(867) = -1.86$, $p < .063$], see Figure 5(c). Overall, these results show that there was a greater improvement in mental conditions that could be considered "acute," such as anxiety and stress, measured by GAD-7 and PSS, which tend to be exacerbated during singular moments. This is in comparison to more "chronic" mental health conditions, such as depression, measured by PHQ-8, which may be slower to respond to digital health interventions.

$$\hat{Y}_{ij} = \exp\left(0.429 + (-0.014) \cdot c_{ij} + (-0.025) \cdot t_{ij} + (-0.012) \cdot (c_{ij} \cdot t_{ij}) + u_{0j}\right) + \varepsilon_{ij}, \quad \varepsilon_{ij} \sim \mathcal{N}(0, 0.227)$$

(4)

$$\hat{Y}_{ij} = \exp\left(0.466 + (-0.001) \cdot c_{ij} + (-0.012) \cdot t_{ij} + (-0.004) \cdot (c_{ij} \cdot t_{ij}) + u_{0j} + \varepsilon_{ij}\right), \quad \varepsilon_{ij} \sim \mathcal{N}(0, 0.085)$$

(5)

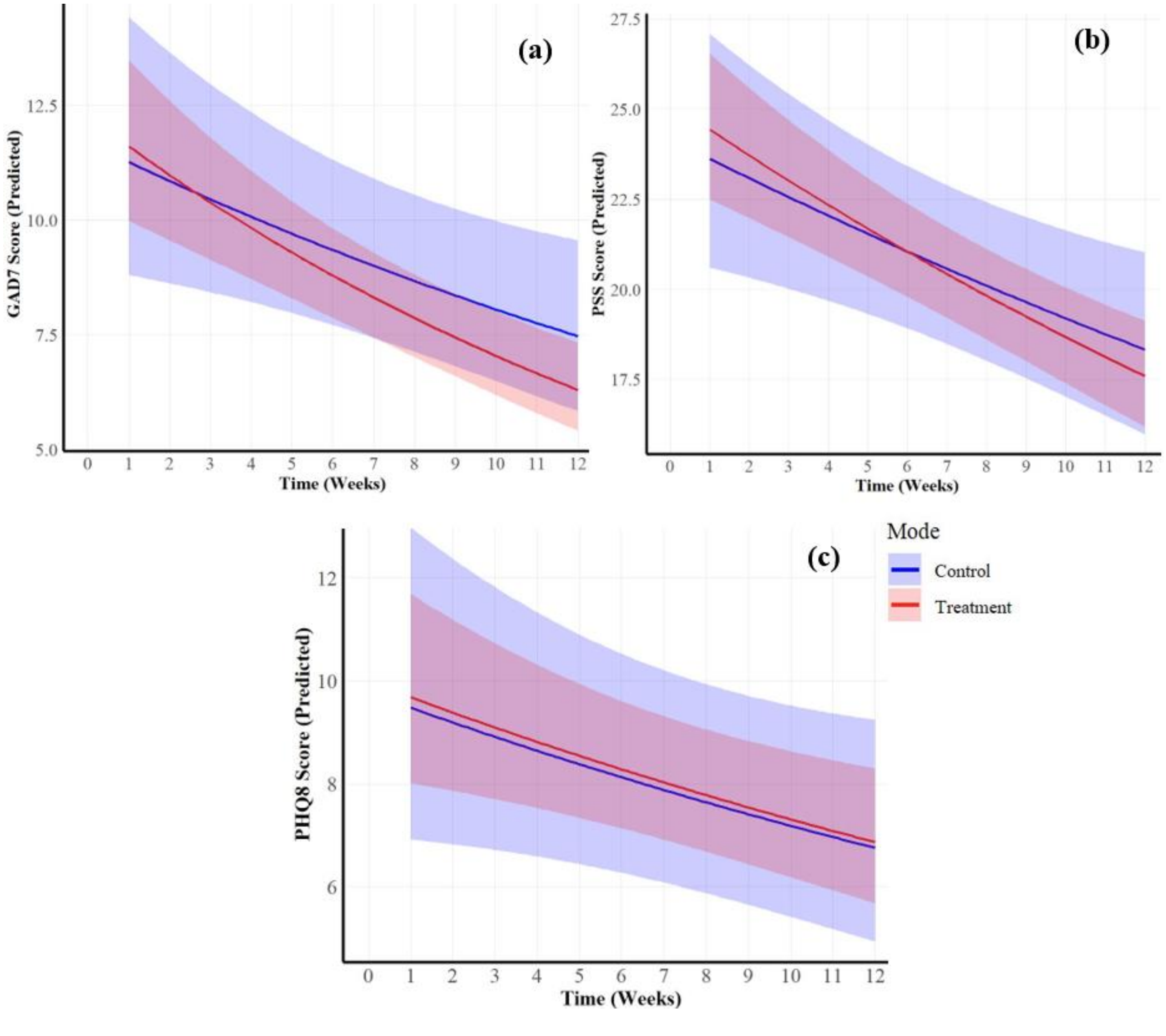

**Figure 5.** Predicted **(a)** GAD7, **(b)** PSS, and **(c)** PHQ8 over time (linear) across the conditions. Shaded areas indicate %95CI.

# DISCUSSION

## Real-Time Stress Reduction through Wearable-Integrated mHealth

This study investigated the efficacy of a mobile health intervention integrating a smartwatch sensor to monitor and manage college students' stress and mental health conditions. Both real-time objective measures, ML-based Moments of Stress (MS), and weekly self-reported subjective measures, including anxiety (GAD-7), perceived stress (PSS), and depressive symptoms (PHQ-8), were collected to ensure the results encompassed a wider breadth of stress and mental health monitoring metrics and were compared between a treatment and control condition. The results demonstrated a significant reduction in MS in the treatment group compared to the control group ($p < 0.001$), aligning with prior findings that real-time interventions via mobile apps and wearables effectively mitigate acute stress responses [16], [23]. However, subjective measures displayed mixed outcomes, which are consistent with earlier studies pointing out the nuanced differences between real-time physiological responses and long-term psychological self-assessments [24], [38], [39]. The discrepancies between objective and subjective outcomes emphasize the complexity highlighted in the literature: physiological improvements may not immediately align with perceived mental health enhancements due to recall bias, temporal delays in symptom awareness, or the fluctuating nature of academic-related stress.

The significant reduction in real-time MS, as modeled by GLMM, suggests that the mobile app effectively reduces acute stress responses by providing real-time interventions and a suite of self-management tools that can be accessed over time. The significant reduction in real-time physiological MS in the treatment group reinforces previous research suggesting that immediate interventions delivered via mobile applications and wearable sensors effectively mitigate acute stress responses [16], [23]. During periods of high stress, the participant was prompted to confirm the validity of the stress moment, bringing attention to said stress and, at the same time, providing suggestions for breathing or other calming exercises to break participants out of a high-stress state. Over time, it is possible that gradual improvements from repeated symptom monitoring and management led to a decreased frequency of moments of stress as students could more easily determine when they were approaching high-stress periods.

**Subjective Mental Health Trends Over Time**

Despite not finding statistical significance between groups for subjective measures of stress and psychological functioning, participants in the treatment condition demonstrated a clinically meaningful improvement in anxiety with an approximate 5.25-point decrease in the modeled GAD-7 total score (from 11.5 to 6.25) between the start and end of the study [39]. This change represents a decrease from a "moderate" level of anxiety (scores 10-14) to a "mild" level of anxiety (scores 5-9) on average throughout the semester. There was also a decrease of about 3.75 points in modeled GAD-7 scores from start to end within the control group (11.25 to 7.5), which is slightly under the clinically meaningful change of 4 points. To be eligible for the study, participants were required to have at least moderate symptoms of anxiety, which made it a clear target for improvement through mHELP interventions. It is likely that the practice of weekly awareness and self-reflection of symptoms within both groups could also help them manage their anxiety. However, the clinically significant change in GAD-7 scores within the treatment group highlights the effectiveness of mobile health interventions for anxiety.

PSS scores declined over time, albeit not statistically significantly, staying in the "moderate" stress range throughout the study. One study estimated the minimal clinically significant absolute change score to be 11 points and the relative change score to be 28% for the PSS score range [38]. This study saw a difference of just over 28% in the modeled PSS scores for the treatment group from the beginning until the end of the study. This suggests that, while not statistically significant, there does appear to be a beneficial effect of the stress management techniques introduced throughout the study on stress for participants. PSS assesses a patient's general level of stress over the past month. While the mHELP app found success in reducing real-time stress, the general level of stress a college student faces during the natural course of a semester may influence their viewpoint of their stress condition.

The absence of significant changes in PHQ-8 scores over time or between groups suggests that the intervention's effects may not have been sufficiently robust to influence depressive symptoms within the study's duration. However, it is essential to note that baseline depression scores for participants in the treatment group were in the "mild" range and below the clinical cut-off score of 10 for clinically significant depression. Therefore, there was little room for meaningful improvement in depression scores. Future directions of this research will consider recruiting participants with higher depression scores before conclusions can be drawn on the efficacy of mHELP for improving depression. Also, since the participants started the study with a moderate level of anxiety and mild depression, it is possible that participants gravitated toward strategies that specifically target stress and anxiety. This highlights the importance of having specific intervention strategies that focus directly on depressive symptoms, such as structured behavioral activation modules or mood-tracking tools. Future research may also consider integrating such components more explicitly into mobile health applications to enhance their efficacy in managing depression within college student populations.

**Divergence and Convergence between Physiological and Self-Reported Outcomes**

On a daily basis, objective real-time MS and on a weekly basis, subjective measures (GAD-7, PHQ-8, and PSS) are different but still complementary. The previous studies underscored that mHealth apps, despite their effectiveness, are not replacements but rather complementary to traditional mental health therapies [22]. This integrative view was evident in our intervention, as mHELP was designed to manage stress independently and offer seamless access to telehealth counseling. Although detailed usage data were unavailable, the potential synergy between these two intervention forms may have further contributed to improvements in stress management, supporting the complementary approach advocated by previous research. MS are based on real-time data from wearable sensors and self-reports, while subjective measures rely on students' perceptions over a longer period. This difference matters because college students face constant academic and social pressures, leading to stress levels that change quickly. The MS data captures these short-term stress responses, while self-reported measures reflect overall feelings over time, which may not always match. In this study, students in the treatment group showed a clear drop in MS, but their self-reported anxiety, depression, and stress decreased more slowly. This suggests that while students may experience less stress at the moment, their overall perception of mental health takes longer to change. Understanding these differences is essential for designing better mental health tools that combine both real-time and long-term data to support students more effectively.

**mHealth as a Complement to Traditional Counseling**

The results of this study are largely consistent with the broader literature on digital mental health interventions (DMHIs) in college populations. In line with prior systematic reviews and meta-analyses, clinically meaningful reductions in anxiety and perceived stress were observed, albeit without statistically significant differences between groups [17]. This finding underscores the potential of mobile, self-guided digital interventions to address acute psychological distress in young adults. These findings also expand upon the field by demonstrating a significant reduction in real-time stress moments, as detected by a machine learning algorithm applied to physiological smartwatch data, a novel measure of mental health that has not been empirically studied. While most DMHI studies rely on self-report outcomes, our inclusion of passive sensing offers novel evidence that digital tools can influence objective, moment-level stress responses in naturalistic settings. Notably, no significant improvements in depressive symptoms were observed, diverging from some prior studies that reported small to moderate effects [40]. This may reflect the slower trajectory of mood-related improvement relative to stress and anxiety, particularly in non-clinical student populations, or the intervention's greater emphasis on stress management and emotion regulation over mood-focused content. Taken together, these findings support the efficacy of DMHIs for managing acute stress and anxiety while highlighting the need for longer-term strategies or more targeted content to address depression. Future research should explore digital health interventions that integrate both real-time physiological feedback and scalable mental health content to optimize impact across symptom domains with greater integration to traditional care models.

Additionally, this study took place over the course of one semester. From the perspective of a student, it is possible that many participants had high levels of excitement and exacerbated symptoms of anxiety, depression, and stress at the start of the semester due to new environments, social and professional changes, or other pressures. These symptoms may slowly resolve throughout the semester alongside the frequent use of the mHELP app, leading to a slow decline in subjective measures. It is also important to note that students in the treatment group had the option to receive a number of telehealth counseling sessions through mHELP's connection to MEND counseling [41], which may have additional therapeutic support, potentially amplifying the intervention's positive effects on anxiety, depression, and stress when applied in tandem with the self-management strategies available in the mHELP app. However, the specific data regarding which students and the number of sessions they opted into is unavailable.

**Limitations and Future Directions**

There were several limitations to the study that may have had an impact on the overall results and should be considered when interpreting the findings. The sample size in this study was increased to account for the high level of attrition and dropout rate associated with mental health studies. A larger sample size generally enhances statistical power and reduces the risk of Type II errors (false negatives). However, when the sample size becomes too large, even very small and potentially meaningless differences may become statistically significant, which can lead to misleading conclusions if practical significance is not considered [42]. An additional limitation of this study is that self-reported mental health assessments, such as the GAD-7, PHQ-8, and PSS, are inherently subject to recall bias, social desirability bias, and individual differences in stress perception, which may explain the lack of significant group differences despite observed downward trends [43]. The study is also somewhat limited in determining the distinction of efficacy between self-management techniques and traditional counseling, as data on the number of telehealth counseling sessions each student used was not available. However, given that access to telehealth counseling was also a feature of the mHELP app, it can be argued that both fit into the category of mobile digital health interventions. Although both subjective and objective measures used in this study have limitations, physiological stress detection may provide a more reliable marker of acute stress, whereas self-reports reflect broader, long-term psychological states. Future studies should address these limitations by incorporating longer intervention periods, personalized recommendations that leverage AI capabilities, and qualitative feedback to better understand user experiences and potential improvements in digital mental health interventions. Additionally, more features that specifically target different mental health, such as depression, should be implemented in future iterations. Further exploring depression, which had inconclusive simple slope analysis, may give more insights into how to improve this digital health intervention.

## CONCLUSION

This study demonstrates that a mobile health intervention integrating smartwatch sensors and real-time feedback can effectively reduce acute stress moments, underscoring both the promise of physiological monitoring and the intervention's effectiveness in real-world mental health management. However, while real-time, momentary self-reports (e.g., during "Moments of Stress" events) aligned with physiological stress detection, traditional weekly self-reported measures—designed to capture broader, chronic symptoms of anxiety and depression—did not show statistically significant between-group differences. This divergence highlights an important distinction between acute, context-sensitive assessments of stress and global or clinical evaluations of psychological well-being. Future interventions should account for this differentiation by integrating personalized, real-time engagement strategies alongside tools designed to track and improve long-term mental health outcomes. Additionally, extended study durations and qualitative feedback may yield richer insights into how individuals internalize and respond to digital health interventions. Overall, these findings advance the growing field of digital mental health by emphasizing the need to align real-time sensing with both immediate and longitudinal psychological support.

**Author Contribution**

A.T., N.S., and M.D. contributed to conceptualization, methodology, data curation, formal analysis, visualization, and writing – original draft. K.P.R., R.K.M., A.M., and C.M. contributed to protocol development, investigation, and writing – review & editing. F.S. was responsible for supervision, project administration, funding acquisition, and writing, review & editing. All authors read and approved the final manuscript.

**Funding**

This work was supported by the President's Excellence Award (X-grant) program at Texas A&M University.

**Declaration of competing interest**

The authors declare that they have no known competing financial interests or personal relationships that could have appeared to influence the work reported in this paper.

**Data Availability**

Data will be made available on request.

**Acknowledgments**

The authors also acknowledge our lab member, Zheng Xiao, who provided support in data collection, app development, and execution.